# Coupling of Spin and Orbital Motion of Electrons in Carbon Nanotubes

F. Kuemmeth[*], S. Ilani[*], D. C. Ralph and P. L. McEuen

*Laboratory of Atomic and Solid State Physics, Department of Physics, Cornell University, Ithaca NY 14853*

[*] *These authors contributed equally to this work*

**Electrons in atoms possess both spin and orbital degrees of freedom. In non-relativistic quantum mechanics, these are independent, resulting in large degeneracies in atomic spectra. However, relativistic effects couple the spin and orbital motion leading to the well-known fine structure in their spectra. The electronic states in defect-free carbon nanotubes (NTs) are widely believed to be four-fold degenerate[1-10], due to independent spin and orbital symmetries, and to also possess electron-hole symmetry[11]. Here we report measurements demonstrating that in clean NTs the spin and orbital motion of electrons are coupled, thereby breaking all of these symmetries. This spin-orbit coupling is directly observed as a splitting of the four-fold degeneracy of a single electron in ultra-clean quantum dots. The coupling favours parallel alignment of the orbital and spin magnetic moments for electrons and anti-parallel alignment for holes. Our measurements are consistent with recent theories[12,13] that predict the existence of spin-orbit coupling in curved graphene and describe it as a spin-dependent topological phase in NTs. Our findings have important implications for spin-based applications in carbon-based systems, entailing new design principles for the realization of qubits in NTs and providing a mechanism for all-electrical control of spins[14] in NTs.**

Carbon-based systems are promising candidates for spin based applications such as spin-qubits[14-19] and spintronics[20-23] as they are believed to have exceptionally long spin coherence times due to weak spin-orbit interactions and the absence of nuclear spin in the $^{12}$C atom. Carbon NTs may play a particularly interesting role in this context because in addition to spin they offer a unique two-fold orbital degree of freedom that can also be used for quantum manipulation. The latter arises from the two equivalent dispersion cones (K and K') in graphene, which lead to doubly-degenerate electronic orbits that encircle the nanotube circumference in a clockwise (CW) and counter-clockwise (CCW) fashion[24] (Fig 1a). Together, the two-fold spin degeneracy and two-fold orbital degeneracy are generally assumed to yield a four-fold-degenerate electronic energy spectrum in clean NTs. Understanding the fundamental symmetries of this spectrum is at the heart of successful manipulation of these quantum degrees of freedom.

A powerful way to probe the symmetries is by confining the carriers to a quantum dot (QD) and applying a magnetic field parallel to the tube axis, $B_{\parallel}$[4,5,8,10,24,25]. The confinement creates bound states and the field interrogates their nature by coupling independently to their spin and orbital moments. In the absence of spin-orbit coupling, such a measurement should yield for a defect-free NT the energy spectrum shown in figure 1b. At $B_{\parallel} = 0$ the NT spectrum should be four-fold degenerate. With increasing $B_{\parallel}$ the spectrum splits into pairs of CCW and CW states (going down and up in energy respectively), each pair having a smaller internal spin splitting. Indications of approximate four-fold degeneracy have been observed in high-field measurements of electron addition spectra[2-10] and inelastic cotunneling[4,10] in nanotube QDs. However, in previous experiments disorder-induced splitting of the orbital degeneracy and electron-electron interactions in multi-electron QDs have masked the intrinsic symmetries at low energies.

In this work we directly measure the intrinsic electronic spectrum by studying a single charge carrier, an electron or a hole, in an ultra-clean carbon nanotube QD. Remarkably, we find that the expected four-fold symmetry and electron-hole symmetry are broken by spin-orbit (SO) coupling, demonstrating that the spin and orbital motion in NTs are not independent degrees of freedom. The observed SO coupling further determines the filling order in the many-electron ground states, giving states quite different from models based purely on electron-electron interactions.

The geometry of our devices is shown in Fig. 1c. A single small-bandgap NT is contacted by source and drain electrodes, and is gated from below by two gates (see methods). When biased, these gates shift the local Fermi energy in the NT thereby accumulating electrons or holes. In this work we use two independent gates to create a QD that is localized either above the left or above the right gate electrode. This is achieved by choosing appropriate combinations of gate voltages that pin the Fermi energy inside the gap on one side of the device while adding carriers to the other side (Fig 1c). Measurement of the linear conductance, $G = dI / dV_{SD}$, through such a dot (Fig. 1e) shows Coulomb blockade peaks that correspond to the addition of individual carriers to the dot, and allows us to identify the first electron and first hole in the dot (see supplementary information for details). Having a single carrier in the dot enables us to study single-particle levels in the absence of electron-electron interactions, and thus to unambiguously identify the presence of spin-orbit coupling. The results reported here were observed in two independent devices and below we present data from one of them.

We probe the quantum states of the NT using tunnelling spectroscopy. The differential conductance through the dot, $G = dI / dV_{SD}$, is measured as a function of gate voltage, $V_g$, and source-drain bias, $V_{sd}$, as the first electron is added to the dot. Figure 2a shows a typical measurement taken at $B_\parallel = 300 \, \text{mT}$. The transition between the Coulomb blockade regions of zero and one electron features distinct resonances that correspond to the ground state ($\alpha$) as well as the excited states ($\beta$, $\gamma$, $\delta$) of the first electron. Their energies can be obtained from a line cut at constant $V_{sd}$ (Fig 2b), by converting the gate voltages into energies (see methods).

The magnetic field dependence of the one-electron states $\alpha$, $\beta$, $\gamma$ and $\delta$ is measured by taking $V_g$ traces such as in Fig. 2b for different values of $B_\parallel$. This is shown in Fig 2c, where we plot $dI / dV_{SD}$ as a function of $V_g$ and $B_\parallel$. The energies of the states $\alpha$ and $\beta$ decrease with increasing $B_\parallel$, hence we identify them as CCW orbital states. The states $\gamma$ and $\delta$ increase in energy and are thus identified as CW orbital states. From the slopes of these resonances with respect to magnetic field we extract an orbital moment of $\mu_{orb} = 1.55$ meV/T and estimate the NT diameter to be $d \approx 5 \, \text{nm}$[24].

A striking difference is observed when we compare the measured excitation spectrum with the one predicted in Fig. 1b: At zero magnetic field the four states in our measurement are not degenerate but rather split into two pairs. To identify the nature of this splitting we note that with increasing magnetic field the energy difference between the states $\alpha$ and $\beta$ increases while the difference between states $\gamma$ and $\delta$ decreases, and both differences are consistent with a g-factor of an electron spin (Figure 2d). This observation allows us to identify unambiguously the spin and orbital composition of each energy level, as shown in the inset of Fig. 2c. At $B_{\parallel} = 0$ the four-fold degeneracy is split into two Kramer doublets – the lower-energy doublet involves states with parallel alignment of orbital and spin magnetic moments, whereas the higher-energy doublet has states with anti-parallel alignment. The zero-field splitting is therefore identified as a spin-orbit splitting, with a value of $\Delta_{SO} = 0.37 \pm 0.02$ meV (extracted from Fig. 2d).

At low fields (Figure 2e) the intersections of states with opposite spin directions (e.g. $\alpha$ and $\gamma$) show simple crossing, whereas states with parallel spin (e.g. $\beta$ and $\gamma$) show avoided crossing, a signature of disorder-induced mixing between CCW and CW orbits ($\Delta_{KK'}$). In previous experiments, the disorder-induced mixing was significantly larger, presumably obscuring the effects of SO coupling. In our measurements, the mixing is small, $\Delta_{KK'} \approx 65\ \mu\text{eV} << \Delta_{SO}$, probably due to smooth electronic confinement, enabling the observation of SO effects. We further demonstrate the intrinsic nature of the effect by measuring identical excitation spectra for QDs formed at different locations along the same NT (Supp. Fig S1).

Next, we show that SO coupling significantly affects the many-body ground states of multiple electrons in a QD. Figure 3a shows the magnetic field dependence of the addition energies for the $N$-electron ground states ($N$=-2 to +4), obtained by measuring the linear conductance as a function of $V_g$ and $B_{\parallel}$. Near zero magnetic field the sign of $dV_g / dB_{\parallel}$ changes every time an electron is added (or removed), indicating that CCW and CW states are filled alternately. Similar addition sequences were explained in the past by repulsive electron-electron interactions driving electrons to occupy different orbits[2-7,9,26] (Fig. 3b). However, in our nanotubes the underlying mechanism is entirely different. Comparing the one-electron excitation spectrum with the two electron-ground state (Fig. 3c), we see that the latter follows exactly the first excited state of the one-electron QD. Specifically, both start with a CW slope at low fields and flip to a CCW slope at the field associated with the SO splitting, $B_{\parallel} \approx 125$ mT. Thus the two-electron ground state is explained entirely by SO coupling (Fig. 3d). Note that below $B_{\parallel} \approx 125$ mT SO favours each of the two electrons to possess parallel orbital and spin moments, forcing them into two different orbital states. Therefore, the two-electron ground state is neither the spin-triplet state predicted by the electron-interaction-based models nor a spin singlet, but rather a Slater determinant of two single-electron states each of which have parallel orbital and spin magnetic moments.

SO effects are commonly assumed to be negligible in carbon-based systems due to the weak atomic SO splitting in carbon $\left( \Delta_{at} = E(^{2}P_{3/2}) - E(^{2}P_{1/2}) \sim 8\ \text{meV} \right)$[27] and its almost perfect suppression in flat graphene[13]. However, recent theories have argued that SO coupling can nevertheless be significant in carbon NTs due to their curvature and cylindrical topology[12,13]. The predicted effect is illustrated in Figure 4a. Consider an electron with a spin moment pointing along the NT axis and orbiting around the NT circumference. The electron occupies the $p_z$ orbitals of the carbon atoms, which are pointing perpendicular to the NT surface. In the rest frame of the electron the underlying $p_z$ orbital revolves around the spin exactly once every rotation, independent of the details of the electron trajectory. In the presence of atomic SO coupling a constant phase accumulates during each rotation, which can therefore be described by a *spin-dependent* topological flux, $S_{\parallel}\phi_{SO}$ passing through the NT cross section ($S_{\parallel} = +1/-1$ for spin moment parallel/antiparallel to the NT axis). This flux modifies the quantization condition of the wavefunction around the circumference:

$$k_{\perp}\pi d \rightarrow k_{\perp}\pi d - 2\pi\, S_{\parallel}\,\phi_{SO}\,/\,\phi_0\,, \qquad (1)$$

where $k_{\perp}$ is the electron's wave-vector in the circumferential direction as measured from the K and K' points, $d$ is the tube diameter and $\phi_0$ is the flux quantum. According to the theory in Ref [12] the flux is given by:

$$\phi_{SO} = \frac{\Delta_{at}}{12\varepsilon_{\pi\sigma}}\left(5 + 3\frac{V_{pp}^{\sigma}}{V_{pp}^{\pi}}\right)\phi_0 \approx 10^{-3}\,\phi_0\,, \qquad (2)$$

where $\varepsilon_{\pi\sigma}$ is the energy splitting of the $\pi$ and $\sigma$ bands in graphene and $V_{pp}^{\sigma}$, $V_{pp}^{\pi}$ are the hopping elements within these bands. This flux does not depend on the geometrical properties of the NT such as its diameter or the shape of its cross section, signifying its topological origin.

Figure 4b illustrates the consequences of the modified quantization conditions for a small-bandgap tube at $B_{\parallel} = 0$. Near each Dirac cone (K and K') there are two quantization lines for the two spin directions (dashed lines). Combining Eq. 1 with the linear dispersion, and including the Aharonov-Bohm flux induced by $B_{\parallel}$, $\phi_{AB} = B_{\parallel}\pi d^2\,/\,4$, and the Zeeman spin coupling, the energies are:

$$E = \pm\hbar v_F\sqrt{k_{\parallel}^{\,2} + k_{\perp}^{\,2}} - \frac{g}{2}\mu_B S_{\parallel}B_{\parallel}\,, \quad k_{\perp} = k_{\perp,0} + \frac{2}{d}\frac{\phi_{AB}}{\phi_0} + S_{\parallel}\frac{2}{d}\frac{\phi_{SO}}{\phi_0} \qquad (3)$$

Here $v_F$ is the Fermi velocity, $k_{\parallel}$ is the wave vector parallel to the NT axis, and $k_{\perp,0} = \pm E_{gap}\,/\,2\hbar v_F$ accounts for the small bandgap, $E_{gap}$, at zero magnetic field (the opposite signs are for the K' and K points). The resulting energy spectrum is schematically shown in Figure 4c, and is in agreement with our measurements. From Eq. (3) we see that the SO energy splitting $\Delta_{SO} = \frac{4\hbar v_F}{d}\frac{\phi_{SO}}{\phi_0}$ (assuming $k_{\parallel} = 0$) is inversely proportional to $d$. Using the estimated diameter of our NT, $d \approx 5\,\mathrm{nm}$, and the measured splitting (Fig. 2d) we obtain the value $\Delta_{SO} \approx 1.9\,\mathrm{meV}\,/\,d[\mathrm{nm}]$, in agreement with the predicted[13] value of $\Delta_{SO} \approx 1.6\,\mathrm{meV}\,/\,d[\mathrm{nm}]$.

An interesting prediction of the theory[12,13,18] is the breaking of electron-hole symmetry. In the absence of SO interactions the low-energy spectrum of a NT exhibits electron-hole symmetry such that each allowed state has a matching state with opposite energy; i.e., the spectrum is symmetric upon reflection about the line $E = 0$. In the presence of SO interactions and an applied magnetic field, Eq. (3) predicts that this symmetry is broken, as is evident from the absence of mirror symmetry around $E = 0$ in the spectrum in Fig. 4c. For $\phi_{SO} > 0$ the theory predicts that in the one-electron ground state the orbital and spin magnetic moments are parallel, whereas in the one-hole ground state they are anti-parallel. This result allows us to test the breaking of electron-hole symmetry experimentally.

The measured excitation spectra for the first hole in the QD (Fig. 4d) clearly shows a SO splitting at $B_{\parallel} = 0$, and a spin g-factor equal to that of the one-electron QD (Fig. 4e). However, in contrast to the one-electron case, here the ground state (α) and the first excited state (β) *converge* with increasing $B_{\parallel}$, implying that the orbital and spin moments are aligned anti-parallel in the one-hole ground state, opposite to the one-electron case. This observation qualitatively confirms the scheme in Figure 4c. We note, however, that the SO splitting observed for the hole $\left( \Delta_{SO} = 0.21 \pm 0.01 \, \mathrm{meV} \right)$ is somewhat smaller than that of the electron, a difference that is not accounted for by current theory. This might result from different confinement lengths (different $k_{\parallel}$ in Eq. 3) or different electric fields (i.e. different $\left| V_g \right|$) for electrons and holes, but current theories predict an effect that is too small to explain this observation.

The existence of SO coupling in carbon nanotubes invalidates several common assumptions about the nature of the electronic states in this system, such as four-fold degeneracy and electron-hole symmetry, and further leads to the existence of entangled spin and orbital multi-electron ground states. Currently, carbon-based systems are considered to be excellent candidates for spin-based applications in part due to the belief that they have weak spin-orbit interactions. Here we showed that this hypothesis is wrong for NTs. Nevertheless, rather than excluding spin-based devices in NTs, our findings may actually promote their feasibility, as long as new design principles are adopted for qubits and spintronic devices, which make use of the strong spin-orbit coupling. This coupling can provide a valuable capability, so far missing in carbon systems: the ability to use electrical gates to manipulate the spin degree of freedom, through its coupling to the orbital electronic wavefunction[14].

**Methods**

Devices were fabricated from degenerately doped silicon-on-insulator wafers, with a 1.5 μm thick device layer on top of a 2 μm buried oxide. Two electrically isolated gate electrodes (Fig. 1d) were patterned from the device layer using dry etching and thermal oxidation (thickness 100 nm). Gate contacts (2/50 nm Ti/Pt), source and drain electrodes (5/25 nm Cr/Pt) as well as catalyst pads were patterned using e-beam lithography. NTs were grown in the last step in order to produce clean devices[8]. Measurements were performed in a $^3$He/$^4$He dilution refrigerator at base temperature (T = 30 mK), using standard lock-in techniques with small excitations (typically $4 - 10 \, \mu$V). The electron temperature extracted from Coulomb peak widths was $100-200$ mK. The conversion from gate voltage to energy is obtained from the bias dependence of the tunnelling resonances, such as in Fig. 2a, and is $\alpha$=0.57 for the first electron (Fig. 2) and $\alpha$=0.58 for the first hole (Fig. 4).

**Acknowledgements**

We thank E. Altman, Y. Gefen, C. L. Henley, Y. Meir, E. Mueller, Y. Oreg, E. I. Rashba, A. Stern and B. Trauzettel for discussions. This work was supported by the NSF through the Center for Nanoscale systems, and by the MARCO Focused Research Center on Materials, Structures and Devices. Sample fabrication was performed at the Cornell node of the National Nanofabrication Users Network, funded by NSF.

**Figures**

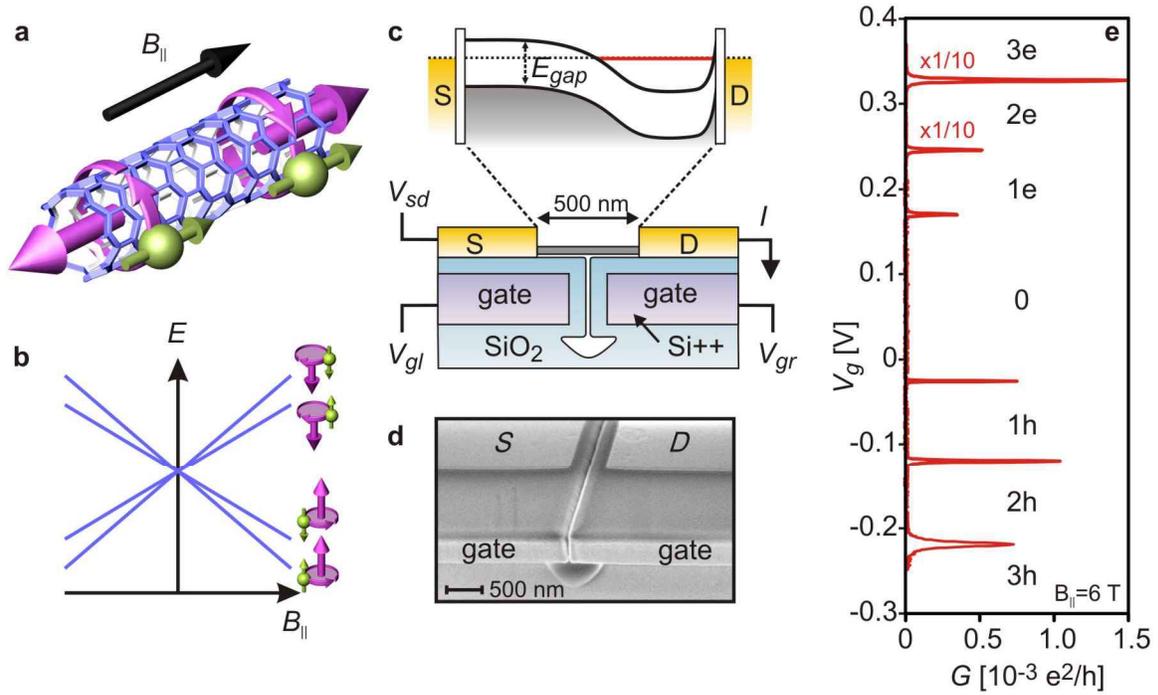

**Figure 1: Few-electron carbon nanotube (NT) quantum dot devices.**

**a**, Electrons confined in a NT segment have quantized energy levels, each four-fold degenerate in the absence of spin-orbit coupling and defect scattering. The purple arrow at the left (right) illustrates the current and magnetic moment arising from clockwise (counter-clockwise) orbital motion around the NT. The green arrows indicate positive moments due to spin. **b**, Expected energy splitting for a defect-free NT in a magnetic field $B_\parallel$ parallel to the NT axis in the absence of spin-orbit coupling: At $B_\parallel = 0\,\text{T}$ all four states are degenerate. With increasing $B_\parallel$ each state shifts according to its orbital and spin magnetic moments, as indicated by purple and green arrows respectively. **c**, Device schematic. A single NT is connected to source and drain contacts, separated by 500 nm, and gated from below by two gate electrodes. The two gate voltages ($V_{gl}$, $V_{gr}$) are used to create a quantum dot localized above the right or left gate electrodes. The energy band diagram is shown for the first case. **d**, Scanning electron micrograph of the device. **e**, The measured linear conductance, $G = dI/dV_{sd}$, as function of gate voltage, $V_g$, for a dot localized above the right gate ($B_\parallel = 6\,T$, $T = 30$ mK). The number of electrons or holes in the dot is indicated. The conductance of the top two peaks is scaled by $1/10$.

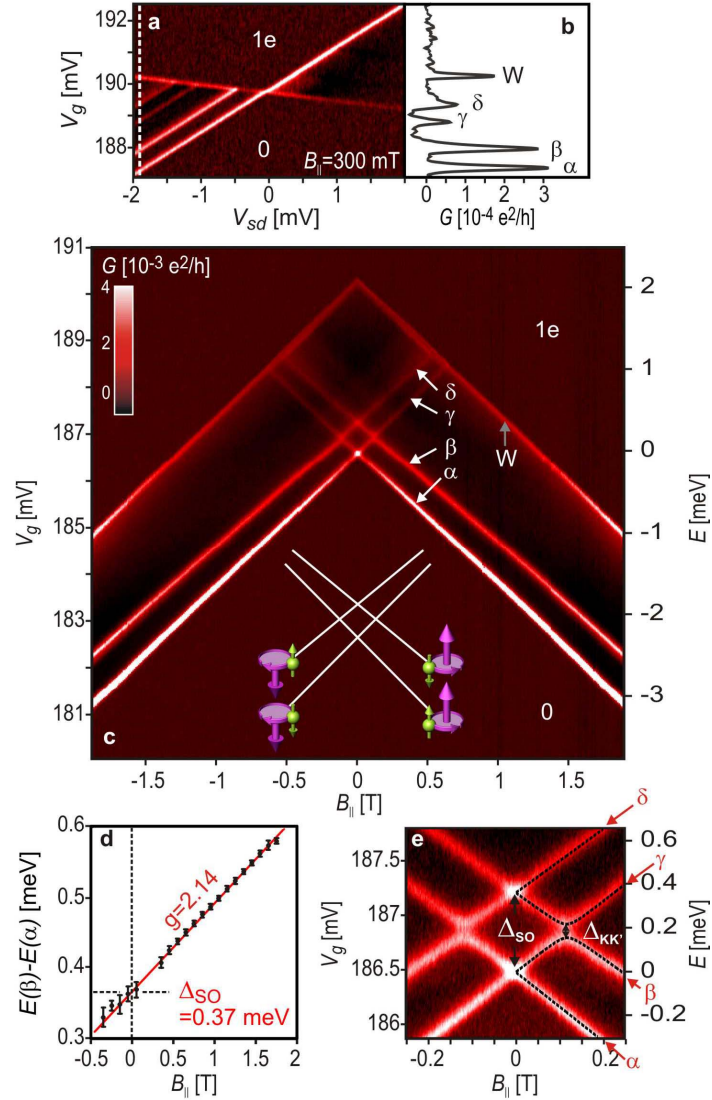

**Figure 2: Excited-state spectroscopy of a single electron in a NT dot**.

**a**, Differential conductance, $G = dI/dV_{sd}$, measured as function of gate voltage, $V_g$, and source-drain bias, $V_{sd}$, at $B_\parallel = 300$ mT, displaying transitions from zero to one electron in the dot. **b**, A line cut at $V_{sd} = -1.9$ mV reveals four energy levels α, β, γ and δ as well as another peak $w$ corresponding to the edge of the one-electron Coulomb diamond. **c**, $G = dI/dV_{sd}$ as a function of $V_g$ and $B_\parallel$ at a constant bias $V_{sd} = -2$ mV. The resonances α, β, γ, δ and $w$ are indicated. The energy scale on the right is determined by scaling $\Delta V_g$ with the lever arm $\alpha = 0.57$ extracted from the slopes in **a**. Inset: Orbital and spin magnetic moments assigned to the observed states. **d**, Extracted energy splitting between the states α and β as a function of $B_\parallel$ (dots). The linear fit (red line) gives a Zeeman splitting with $g = 2.14 \pm 0.1$, and a zero-field splitting of $\Delta_{SO} = 0.37 \pm 0.02$ meV. (s.d. error bars) **e**, Magnified view of panel **c** showing the zero-field splitting due to SO interaction ($\Delta_{SO}$) as well as finite-field anti-crossing due to $K$-$K'$ mixing ($\Delta_{KK'}$). Dashed lines show the calculated spectrum using $\Delta_{KK'} = 65 \mu$eV .

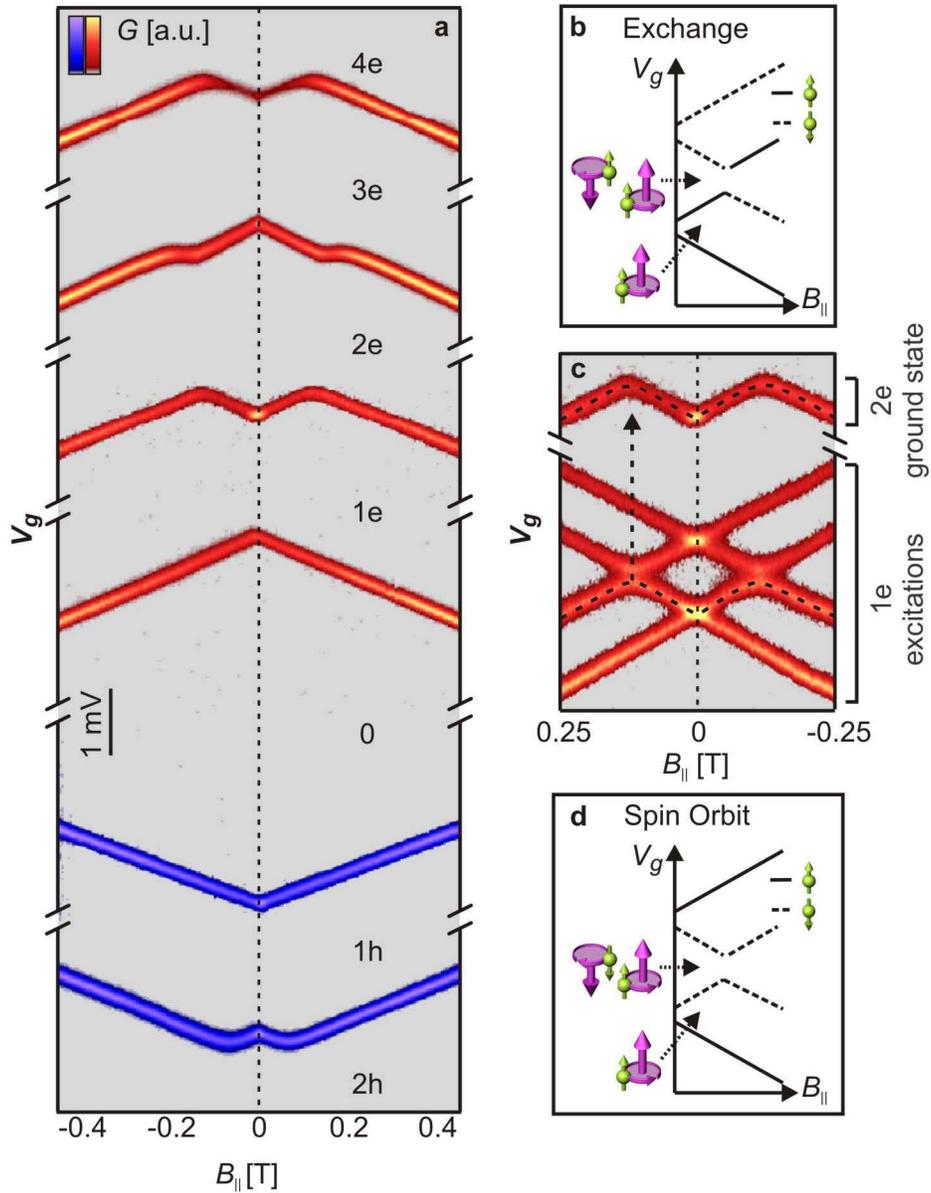

**Figure 3: The many-electron ground states and their explanation by spin-orbit interaction.**

**a**, $G = dI / dV_{sd}$, measured as a function of gate voltage, $V_g$, and magnetic field, $B_{\parallel}$, showing Coulomb blockade peaks (carrier addition spectra) for the first four electrons and the first two holes (data are offset in $V_g$ for clarity). **b**, Incorrect interpretation of the addition spectrum shown in **a** using a model employing exchange interactions between electrons. Dashed/solid lines represent addition of down/up spin moments . The two-electron ground state at low fields, indicated at the left, is a spin triplet. **c**, Comparison of the measured two-electron addition energy from **a** with the one-electron excitation spectrum from Fig. 2**e**. **d**, Schematic explanation of the data in **a** using electronic states with spin-orbit coupling: The two-electron ground state at low fields, indicated on the left, is neither a spin-singlet nor a spin-triplet state.

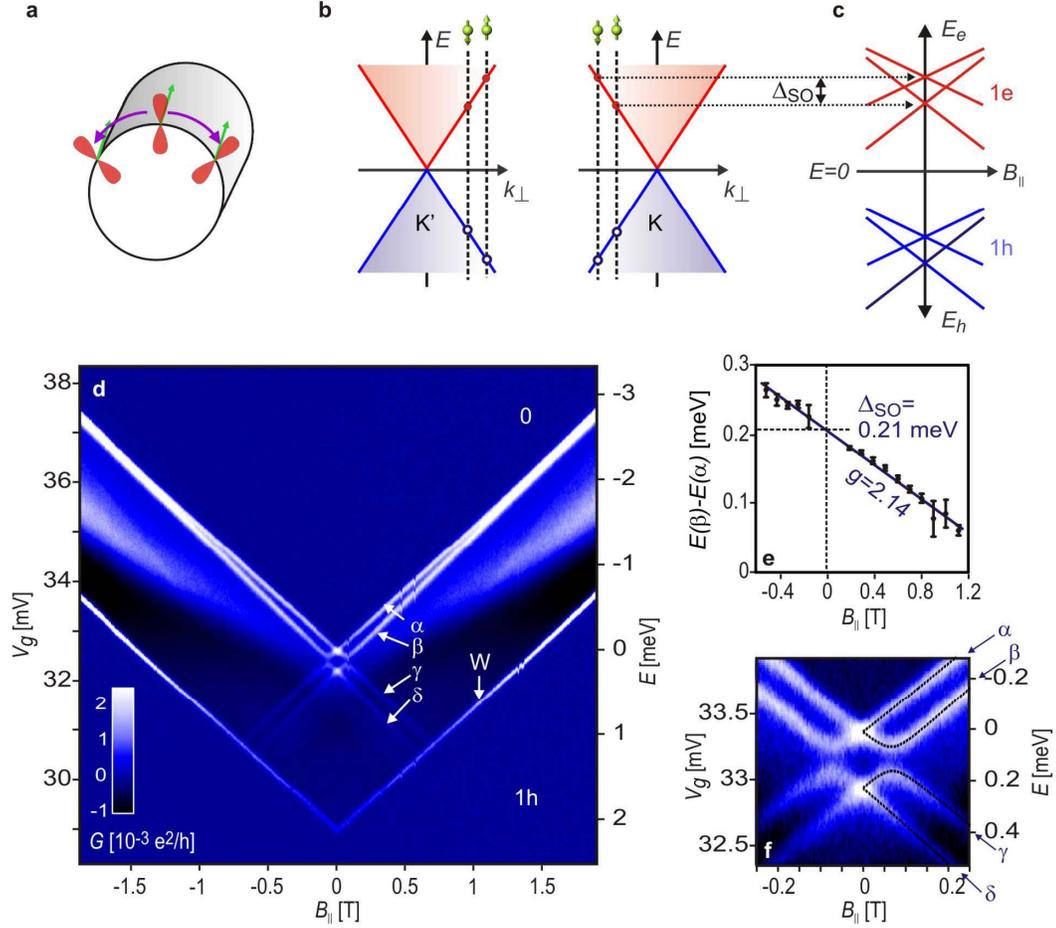

**Figure 4: Theoretical model for SO interaction in nanotubes and the energy level spectroscopy of a single hole**.

**a**, Schematic of an electron with spin parallel to the NT axis revolving around the NT circumference. The carbon $p_z$ orbitals (red) are perpendicular to the surface. In the rest frame of the electron, the $p_z$ orbital rotates around the spin. **b**, Allowed electron and hole energies (red and blue circles) at $B_\parallel = 0$ for a small-bandgap NT with SO interaction. The states are derived by cutting the Dirac cones ($K$ and $K'$) with spin dependent quantization lines (dashed lines). The allowed $k_\perp$-vectors differ for up and down electron spin moments. **c**, Calculated energy levels for an electron (red lines) and a hole (blue lines) as a function of $B_\parallel$. The four distinct slopes arise from the orbital and spin Zeeman shifts. **d**, $G = dI/dV_{sd}$ as a function of $V_g$ and $B_\parallel$ at a constant bias $V_{sd} = -2$ mV. The resonances labelled α, β, γ, δ and $w$ arise from tunnelling of holes onto the dot and therefore the energy scale points opposite to $V_g$. The ground state (α) and first excited state (β) cross at $B_\parallel \approx 1.5$ T. **e**, Extracted energy splitting between the states α and β as a function of $B_\parallel$ (dots). The linear fit (blue line) gives a Zeeman splitting with $g = 2.14 \pm 0.1$, and a zero-field splitting of $\Delta_{SO} = 0.21 \pm 0.01$ meV. (s.d. error bars) **f**, Magnified view of the level crossings in **d** and a model calculation using $\Delta_{KK'} = 0.1$ meV (dashed lines).

## SUPPLEMENTARY INFORMATION

**Identification of the first electron and the first hole**

In this paper we measure the energy spectrum of a single charge carrier, an electron or a hole, in a nanotube (NT) quantum dot (QD). Having a single carrier is central to our experiment as it allows us to avoid electronic interactions between carriers and to unambiguously identify the effects of spin-orbit interactions. Here we explain in more detail how we determine that there is a single carrier in the dot.

We start by identifying the transition from electrons to holes in the addition spectrum with the Coulomb valley labeled "0" in figure 1e, based on two observations: First, this valley is significantly larger than all other valleys, reflecting the added contribution of the bandgap to the addition energy (quantitative analysis below). Second, electrons and holes can be distinguished by their response to magnetic field (Fig. 3a). At large fields, such that the orbital coupling dominates over the level spacing in the dot, electrons and holes rotate in opposite directions around the NT circumference. This leads to opposite signs of $dV_g / dB_\parallel$ for electrons and holes (figure 3a, $B_\parallel$>200 mT). Note that this also means that the energy gap decreases with increasing magnetic field. To confirm that the first Coulomb valleys on the electron and hole sides correspond to the first electron and first hole in the dot it is enough to show that we observe all the charge states in the transport measurements and that there are no non-conducting charge states within the $0^{th}$ Coulomb valley. By applying a high magnetic field we can reduce the size of the $0^{th}$ Coulomb valley such that it doesn't allow for even a single additional charge state in the gap, thus confirming that the first Coulomb peaks in the electron and hole sides correspond to the addition of the *first* electron and *first* hole to the QD.

To analyze this quantitatively, we determine all the parameters of the QD directly from non-linear transport data, similar to those in fig. 2a. Specifically, the charging energies and level spacings between particle-in-a-box longitudinal modes for the first electron and first hole are $U_e = 19\,\text{meV}$, $U_h = 25\,\text{meV}$ and $\Delta_e = 8\,\text{meV}$, $\Delta_h = 11\,\text{meV}$ (see more details below). We estimate the band gap at 6 Tesla by subtracting the average charging energy and average level spacing for electrons and holes from the $0^{th}$ Coulomb valley (55 meV), and obtain $E_{gap} = 24 \pm 2\,\text{meV}$ at 6 Tesla. At the highest field in our measurements (9 Tesla) the size of the energy gap is smaller than the charging energies of either the electron or the hole dot, excluding the possibility of hidden charge states inside the $0^{th}$ Coulomb valley.

**Effects of higher particle-in-a-box longitudinal modes**

The one-electron and one-hole excitation spectra presented in this paper correspond to the lowest quantized longitudinal mode. Quantized states of other longitudinal modes do not appear in the data presented in Figure 2 and Figure 4 because of their higher energy. We verified this by measuring excitation spectra at source drain voltages larger than in Figure 2a, and identifying longitudinal modes by their dependence on the length of the quantum dot. The level spacing extracted for a dot that is extended over both right and left gate electrodes is ~4 meV, and it increases continuously to >8 meV as the dot becomes localized either above the right or above the left gate electrode (using appropriate gate voltages). The latter corresponds to a confinement length smaller then 200 nm and is consistent with the lithographic dimensions given in Figure 1c. Therefore, higher longitudinal modes were ignored in the discussion of the one-electron and one-hole quantum dot.

The situation is different for the top two Coulomb peaks presented in Figure 3a, which involve three and four electrons in the quantum dot. Because of the increased size of the quantum dot at those charge states, the level spacing is reduced and higher modes become occupied already at ~200 mT. At $B_\parallel$>200 mT it is favorable for all electrons to orbit in counterclockwise direction, thereby aligning their orbital magnetic moments parallel to the external magnetic field. This explains why the 3e and 4e addition spectra shown in Fig. 3a deviate for $B_\parallel$>200 mT from the 3[rd] and 4[th] excitations of the one-electron QD.

**Comparison between the QD above the left and right gate electrodes**

The ability to localize the QD on two physically different segments of the same nanotube (Figure S1) helps us determine whether the excitations observed at low energies depend on local properties in the nanotube such as localized disorder or specific properties of the source or drain electrodes. Figure S1c shows the non-linear conductance measured for a QD localized above the right gate electrode. The measurement is at finite magnetic field ($B_\parallel$=300 mT) allowing to resolve the four quantum states in the dot. The individual states are visible at negative bias and are absent at positive bias, indicating that the coupling of the QD to the source is much weaker than to the drain, as expected from its location. Accordingly, the capacitance between the QD and the source is smaller than the capacitance between the QD and the drain ($C_s = 1/1.8\ C_d$, extracted from the slopes of the resonances in panel c). The energies of the quantum states in the dot extracted from this measurement are 0, 0.40, 0.98 and 1.33 meV (±5%). Figure S1d shows a similar measurement for a QD localized above the left gate electrode. As expected, here the tunnel coupling to the drain is weaker than that to the source, and hence the quantum states are probed by tunneling-in from the drain electrode (i.e. resonances appear at $V_{sd}$>0). The capacitance ratio ($C_s = 2.2\ C_d$) is reversed compared to that in panel c. Most importantly, the excitations in panel c and d have identical energies up to an experimental uncertainty of ±5%. Similar measurements at other magnetic fields have also shown that the QD excitation energies do not depend on whether it is localized above the right or left gate electrode, demonstrating that the observed excitations are an intrinsic property of the nanotube.

**Spin-orbit coupling vs. K-K' scattering**

The four-fold degeneracy in NTs can be broken by extrinsic sources such as disorder, or by the intrinsic spin-orbit coupling. Disorder breaks the orbital symmetry of NTs in a trivial way and leaves doubly-degenerate spin states as in any other confined system with low symmetry. Spin-orbit coupling, on the other hand, breaks the degeneracy by coupling the orbital and spin degrees of freedom in parallel or anti-parallel configurations. Figure S2 shows how the theoretical four-fold energy spectrum for a single electron (Fig. 1b) changes in the presence of disorder (Fig. S2a) or spin-orbit coupling (Fig. S2b). In both cases the spectrum is split at $B_\parallel = 0$ into two Kramer doublets, but the nature of the new eigenstates is entirely different. In the case of disorder, the splitting results from mixing wavefunctions which revolve in opposite directions around the NT circumference, and hence the new eigenstates lack a definite sense of rotation around the circumference. Orbital angular momentum ceases to be a good quantum number for these states as is apparent from the fact that they have no coupling to the field ($dE/dB_\parallel = 0$ at zero field, ignoring the spin Zeeman coupling). In the presence of spin-orbit coupling, the components of angular momentum and spin parallel to the NT axis remain good quantum numbers. This is readily seen from the fact that the slopes in magnetic field, $dE/dB_\parallel$, remain finite even at zero magnetic field. SO interactions thus create non-trivial states in which the spin and orbital degrees of freedom are tied together.

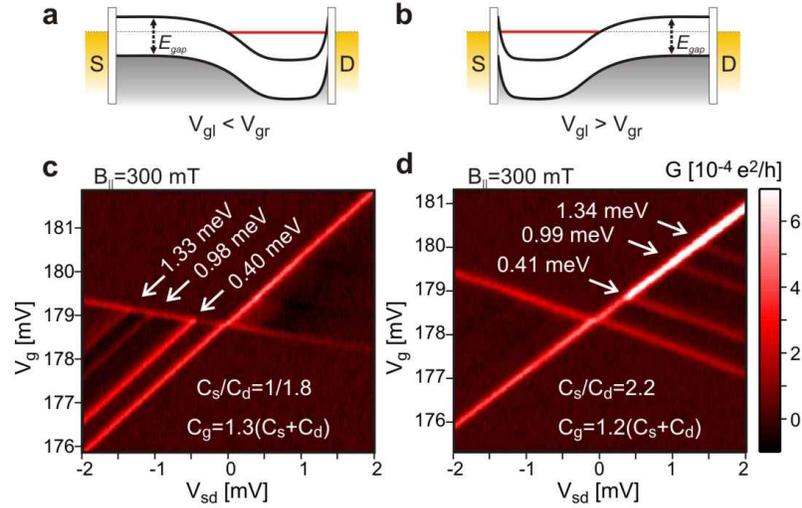

**Figure S1: Independence of the one-electron excitation energies on the QD location.**
**a,** Schematic band diagram for a one-electron QD formed above the right gate electrode. Here the longer barrier on the left side leads to a weaker tunnel coupling to the source electrode. **b,** Same for a QD formed above the left gate electrode (**c**) Differential conductance, $G = dI/dV_{sd}$, measured as function of gate voltage, $V_g$, and source-drain bias, $V_{sd}$, at $B_\parallel = 300$ mT for the transition from zero to one electron for a dot localized above the right gate electrode. The energies of the one-electron excitations that appear at negative $V_{sd}$ are labeled. Also shown are the ratios of the capacitances between the QD and source ($C_s$), drain ($C_d$) and gate ($C_g$) electrodes, extracted from the slopes of the resonances. **d,** Same for a QD localized above the left gate electrode.

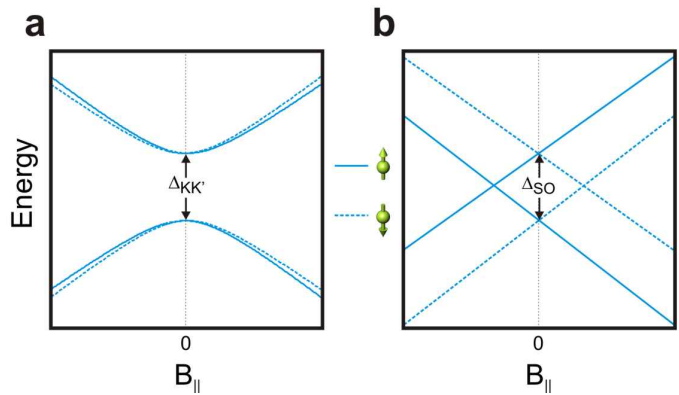

**Figure S2: Breaking of four-fold degeneracy: spin-orbit coupling vs. KK' scattering.**
**a,** The calculated one-electron spectrum as a function of parallel magnetic field in the presence of disorder-induced K-K' scattering and the absence of spin-orbit coupling. Dashed and solid lines correspond to spin moment down and up. **b,** Same, but with spin-orbit coupling and without disorder-induced K-K' scattering.